\documentclass[reprint,
superscriptaddress,
 amsmath,amssymb,
 aps,
]{revtex4-2}

\usepackage{graphicx}
\usepackage{xfrac}
\usepackage{dcolumn}
\usepackage{bm}
\usepackage{xcolor}

\newcommand{\angstrom}{\textup{\AA}}

\begin{document}

\preprint{APS/123-QED}

\title{Local electronic excitations by slow light ions in tungsten}

\author{Evgeniia Ponomareva}%
 \email{evgeniia.a.ponomareva@aalto.fi}
 \affiliation{%
Department of Applied Physics, Aalto University, P.O. Box 11100, 00076, Finland
}%
\author{Eduardo Pitthan}
\author{Radek Holeňák}
\author{Jila Shams-Latifi}
\affiliation{%
Department of Physics and Astronomy, Ångström Laboratory, Uppsala University, Box 516, SE-751 20 Uppsala, Sweden
}%

\author{Glen Pádraig Kiely}
 \affiliation{%
Department of Applied Physics, Aalto University, P.O. Box 11100, 00076, Finland
}%
\author{Daniel Primetzhofer}
\affiliation{%
Department of Physics and Astronomy, Ångström Laboratory, Uppsala University, Box 516, SE-751 20 Uppsala, Sweden
}%
\author{Andrea Sand}
 \affiliation{%
Department of Applied Physics, Aalto University, P.O. Box 11100, 00076, Finland
}%

\date{\today}

\begin{abstract} Accurately predicting the electronic energy deposition of ions in materials is an important challenge in both fundamental and applied research. While employing ab initio simulations to investigate electronic stopping of ions in matter holds promise, its combined use with experimental measurements paves the way for obtaining reliable data. In this paper, we present a collaborative study using real-time time-dependent density functional theory and experimental methods to determine the electronic stopping power of hydrogen and helium ions in tungsten, a primary candidate material for future nuclear fusion devices. While calculated stopping powers in hyperchanneling trajectories are significantly lower than the experimental data, off-center and random geometries demonstrate a better agreement. We show that the deviation from velocity proportionality for both projectiles traversing the hyperchanneling directions can be explained through the existence of a threshold velocity leading to the activation of semicore states. Additionally, we analyse the pseudopotential and the trajectory dependence of computed electronic energy losses. It is demonstrated that the role of including inner-shell electrons varies depending on the velocity range. While these states play a crucial role at high projectile velocities by introducing additional dissipation channels, their impact diminishes in the low-velocity range. Finally, we introduce a simple expression that links electronic energy losses in different trajectories to local electron density, and we show that utilizing this formula allows for quite accurate predictions of stopping powers around the Bragg peak.

\end{abstract}

\maketitle


\section{Introduction}

Understanding the complex processes of radiation damage is of utmost importance in the development of robust materials for fusion applications \cite{brezinsek2017plasma}. These materials must endure prolonged exposure to high-energy particles, intense radiation, and extreme thermal loads while maintaining their structural integrity and functional properties \cite{linke2019challenges}. Gaining insight into the interactions between fusion-relevant materials and the ions present in the plasma is pivotal for designing and optimizing the performance of these materials.

In particular, the energy loss of ions to the target electrons, generally referred to as electronic stopping, plays a critical role in determining the response of materials to irradiation. For cases where interactions with nuclei can be disregarded, it is expressed as the negative derivative of the projectile's kinetic energy with respect to the distance traveled. Although studies of electronic energy losses exist for fusion-relevant materials, the low-velocity data is generally scarce or completely absent. In fusion reactors, the low-velocity regime is seen in the late stages of ion irradiation events, whereby non-adiabatic ion-electron coupling plays a significant role, and in the sputtering of plasma-facing materials by incident plasma particles.  Thus, the understanding of the energy transfer process in the full range from low to high projectile velocities is of great importance to model the response of structural materials to the demanding operating conditions of a fusion reactor.

Many classical and semi-classical analytical models were developed to study electronic stopping. Early works were based on empirical models, such as the Bethe-Bloch formula accounting for Coulomb collisions of projectiles with classical atomic electrons \cite{bethe1930theorie}. This approach provides a simple but rather inaccurate description of the stopping power, diverging in the low-velocity limit. Further corrections were proposed by Fermi and Teller who found the expression linear in the particle velocity using the free-electron gas (FEG) model \cite{fermi1947capture}, and then by Lindhard who introduced the electronic system response through a frequency- and wavevector-dependent dielectric constant \cite{lindhard1954properties}. However, all analytically derived models lacked structural information as a target atomic system was represented as an amorphous medium. As the requirement for a nonlinear, first-principles approach became evident, Density Functional Theory (DFT) found active application in the field of stopping power research. To incorporate the time-dependent nature of electrons and enable simulations of excited states, real-time Time-Dependent Density Functional Theory (rt-TDDFT) emerged. This approach has proven effective in investigating electronic stopping phenomena in diverse materials such as metals, semiconductors, and insulators, shedding light on the mechanisms involved in the energy transfer between ions and electrons \cite{schleife2015accurate,lee2020multiscale,quashie2016electronic, quashie2021directional,li2019effect, li2021first, kononov2021anomalous}.

In particular, many of these studies showed that the electronic stopping power can vary significantly depending on the trajectory of the projectile. Specifically, when particles follow crystal lattice directions, known as channeling trajectories, they tend to exhibit reduced stopping powers due to a lower electron density along their path. Conversely, trajectories considered random result in higher stopping powers due to the closer impacts with target atoms. This difference, however, tends to diminish for low-energy protons (tens of keV), yet becomes even more pronounced for heavier projectiles \cite{quashie2016electronic, quashie2021directional}. In \cite{lohmann2020disparate}, for the first time, H and He data were compared experimentally revealing a similar trend for Si. When traversing random directions with high velocity, both projectiles might encounter close collisions and, subsequently, excite target core electrons, which essentially leads to a pronounced trajectory dependence of the stopping power around the Bragg peak (stopping curve maximum). In a low-velocity regime, core electron excitation becomes less probable, therefore, the environment dependence observed for heavier projectiles should result from other mechanisms. Additionally, within this low-energy spectrum, distinct threshold features have been observed in different materials, particularly deviations from linearity in the velocity dependence of electronic stopping \cite{quashie2016electronic, li2019effect, li2021first, kononov2021anomalous}. These studies often limit analysis to the case of protons, lacking comparison with higher-Z elements. Consequently, this results in a rather incomplete picture of the complex interplay between the projectile electronic structure, its velocity, and trajectory it travels through in a target material.

Additionally, despite the extensive research conducted in the field of electronic stopping, the integration of both experimental measurements and ab-initio simulations remains largely unexplored or limited in scope. This gap poses a significant problem as it hampers the development of accurate and reliable models for predicting energy deposition and material damage. To address this challenge, we propose a synergistic approach that combines ab-initio and experimental research on the electronic stopping of light ions in fusion-relevant material. Ab-initio simulations offer a powerful tool to explore the atomic-scale dynamics, while experimental measurements provide crucial validation. 

Tungsten was chosen as our study target material due to its vital role in fusion research. As a high-Z element with remarkable thermal and mechanical properties, it stands as a prime candidate for plasma-facing components. This work contributes to both the understanding of tungsten behavior under fusion conditions but also to the broader knowledge of material response to extreme irradiation environments.

\section{METHODS}

\subsection{Computational methods}

\subsubsection{Simulation setup}

To carry out rt-TDDFT calculations of electronic stopping we use the open-source code Qb@ll \cite{schleife2014quantum, draeger2017massively} (a time-dependent extension of the DFT code Qbox \cite{gygi2008architecture}) with a plane-wave-pseudopotential implementation. 

We first perform ground-state density simulations for the atomic system with a projectile at rest, placed in the initial coordinate position that will serve as a starting point for its propagation. To describe the exchange-correlation potential, we use the adiabatic local density approximation (ALDA) as a working approximation for interactions of ions in metallic targets \cite{correa2018calculating}. The plane-wave cutoff energy is chosen based on a convergence test and is set to 150 Ry. 

The system, converged after the ground state DFT simulations, is used as an input for time-dependent dynamics, where the projectile is provided with a constant velocity while the target atoms are kept stationary. This approach was shown to be very useful since it allows analyzing the energy loss of the projectile through the total energy increase during the particle propagation \cite{schleife2012plane, schleife2015accurate, correa2018calculating}. Time-dependent Kohn-Sham equations are integrated using a fourth-order Runge-Kutta scheme with a maximum time step of 0.01 a.u. (0.24 as). 

To investigate the impact of semi-core states, we employ norm-conserving pseudopotentials with varying numbers of electrons. Consequently, we compare tungsten pseudopotentials with 12 and 20 explicit electrons, denoted as W12 and W20 in the following sections.

We do not define the charge state of the projectiles a priori, but rather use neutral H and He. Building upon the findings presented in \cite{schleife2015accurate}, our comparison between neutral and charged particles did not uncover any perceptible differences in the electronic stopping calculation results across the entire velocity range considered in this work.

To investigate the directional dependence, we explore three types of trajectories: (i) hyperchanneling, (ii) off-centered channeling, and (iii) random. In case (i), the projectile moves in a perfectly symmetrical direction between the atomic rows of the lattice. This direction samples the minimum electron density values and experiences fewer electronic interactions. In case (ii), the projectile is forced to move along a channel in close proximity to the target atoms. As an example of this case, we consider a particle shifted by $a_0/4$ from the central position towards the nearest atomic row, where $a_0$ represents the lattice constant. The selection of random trajectories (iii) will be described in greater detail in the part 3 of this subsection.

For all combinations of projectiles, target atoms, and trajectories we use a 108-atom supercell constructed by 6 $\times$ 3 $\times$ 3 cubic unit bcc cells and a lattice constant of 3.16 $\angstrom$. 

\subsubsection{Data analysis}

\begin{figure}[t!]
\centering\includegraphics[width=8.7 cm]{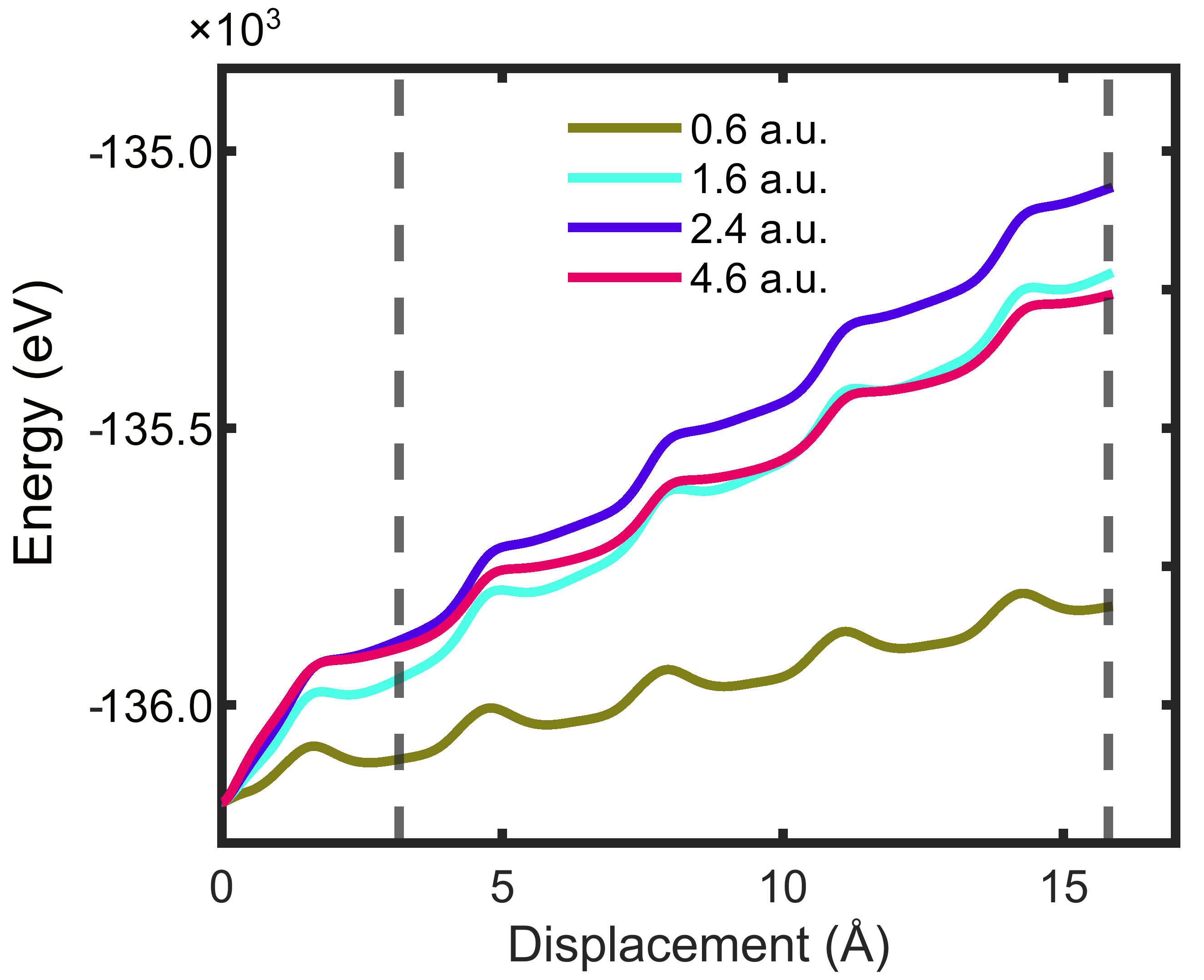}
\caption{Total electronic energy vs displacement of He channeling in tungsten with different velocity from 0.6 to 4.6 a.u. The part of the curve between two vertical dashed lines is used for linear regression fit and the following electronic stopping value extraction. }
\label{fig1}
\end{figure}

The total energy increase as a function of projectile displacement for different velocities is depicted in Fig. \ref{fig1}. To determine the electronic stopping, we utilize the definition of stopping power and perform a linear regression fit on these curves. The slope coefficient of this fit corresponds to the electronic stopping. In order to obtain accurate results, we focus on the periodic part of the energy curve, excluding the initial transient region (indicated by dashed vertical lines in Fig. \ref{fig1}). This transient phase arises due to the abrupt perturbation introduced by the moving projectile and represents an unsteady component of the dynamics that should be disregarded during the analysis.

\subsubsection{Random trajectories pre-selection}

Previous studies have demonstrated that using channeling trajectories for electronic stopping power calculations is not sufficient to accurately capture all possible energy loss pathways between the ion and target atoms \cite{quashie2016electronic,quashie2018electronic,kononov2021anomalous}. Instead, increased accuracy can be achieved using proper sampling of different impact parameters. This can be done in several ways, for instance, in \cite{quashie2016electronic} the random trajectories were chosen on the basis of visual inspection of the supercell and via using the normalized version of $[1,\phi,\phi_2]$, where $\phi$ is the golden ratio. To increase the averaging accuracy these two velocity directions can be combined with different sets of the projectile initial coordinates \cite{quashie2018electronic}. Alternatively, a standard pseudo-random number generator was used in \cite{lee2018electronic} to generate a random direction for projectile propagation. Although it is true that all possible impact parameters will be sampled for a sufficiently long incommensurate trajectory, this approach is computationally inefficient. The generally employed nm-long trajectories do not lead to a satisfactory correspondence of TDDFT electronic stopping with the experiment (e.g. under- or overestimation of the experimental data across the whole velocity range in the previously cited works). 

To tackle this problem, Gu et al. proposed an algorithm \cite{gu2020efficient} for efficient pre-sampling of representative random trajectories. It is based on the correlation between the energy losses of the projectile and the electron density encountered during the propagation. By utilizing the geometric characteristics of the target medium, the algorithm selects a set of short (TDDFT cell-size) trajectories, enabling a rapid convergence towards the computed averaged electronic stopping value using TDDFT.

\begin{figure}[b!]
\centering\includegraphics[width=8.7 cm]{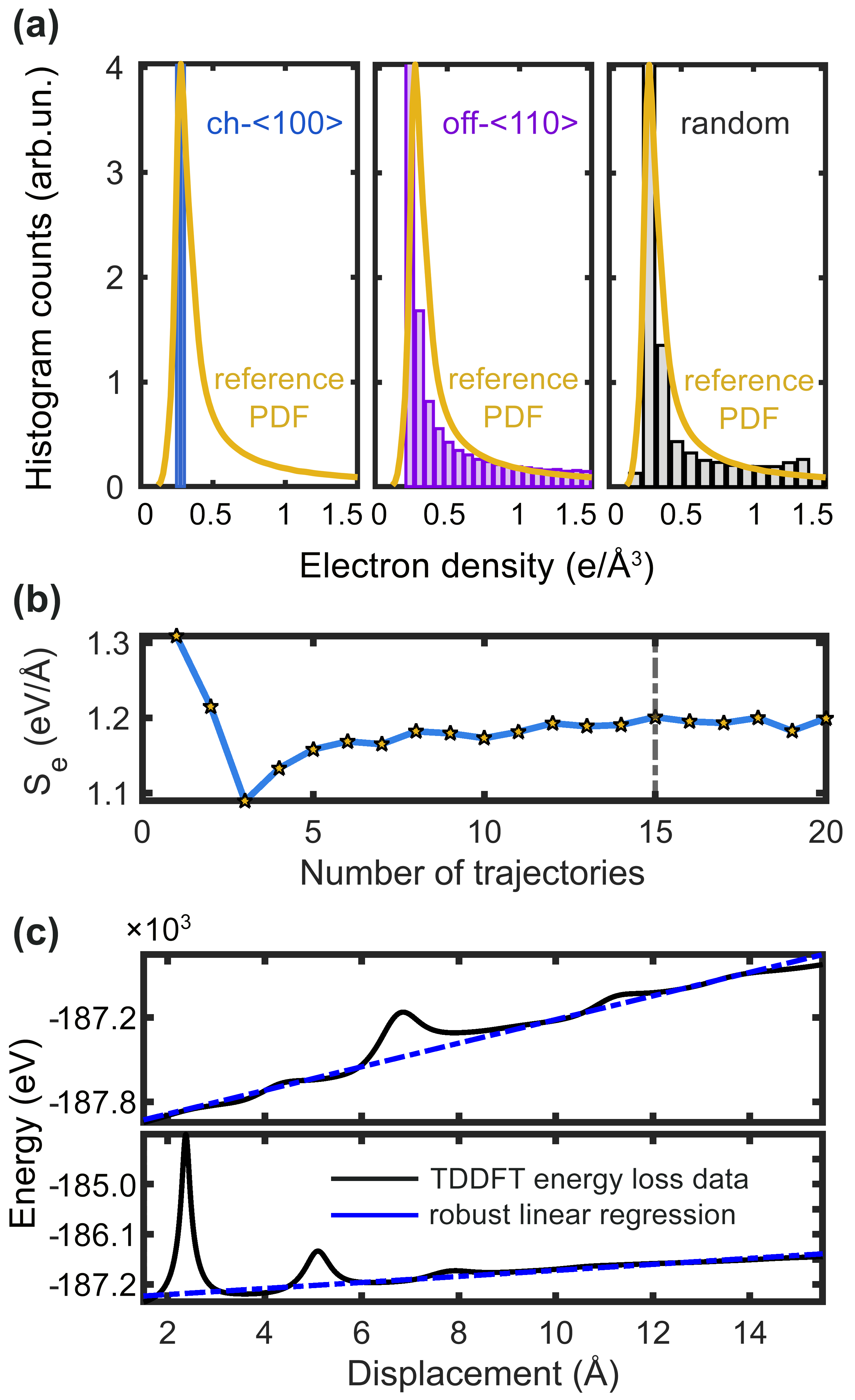}
\caption{Random trajectories pre-selection for TDDFT simulations. (a) Comparison of the reference electron density probability distribution (yellow solid line) with the density distribution along $\langle$100$\rangle$ hyperchanneling (blue bars), $\langle$110$\rangle$ off-centered channeling, and random trajectories. (b) Convergence of the electronic stopping value with respect to the number of optimal trajectories. (c) Examples of energy loss curves for random trajectories (black) overlapped with a bisquare robust fitting (blue).}
\label{fig2}
\end{figure}

In our study, we applied this approach with one primary modification. Rather than utilizing the normalized probability distribution function (PDF) based on the distance between the projectile and the nuclei of the target, we employed a PDF derived from the average electron density computed from the TDDFT cube density file. The primary objective remains unchanged: to closely emulate the reference average density PDF by employing a small number of PDFs obtained from individual short trajectories. Thus, in Fig. \ref{fig2}a we show how different representative trajectories match the reference PDF. The channeling projectile (blue bars) only captures a small fraction of the overall density distribution, while the density along the pre-selected random trajectory (grey bars) appear to be a good representation of the reference PDF. We additionally provide here the data for $\langle$110$\rangle$ off-centered channel (violet bars), since it will serve as a reference when analyzing the corresponding stopping curve in Fig. \ref{fig5}.

As depicted in Fig. \ref{fig2}b, we achieved convergence of the electronic stopping value with just 15 short pre-sampled trajectories for the tungsten lattice. Additionally, Fig. \ref{fig2}c showcases two examples of the resulting energy loss curves. To determine the electronic stopping value for random trajectories, we utilized a bisquare weights robust least-squares fitting method instead of a simpler linear one used to analyze periodic directions. This adjustment leads to a significant reduction in the root-mean-square error (RMSE), with improvements of up to 30-fold observed for the random trajectories.

\subsection{Experimental measurements}

\subsubsection{Experimental setup}

For the experimental measurements, a magnetron sputter-deposited thin film from a tungsten target on a carbon substrate of nominal thickness of 19.25 nm (assuming a nominal W density of 19.35 g/cm$^3$) was used as target. Details of the deposition and film characterization can be found elsewhere \cite{shams2023experimental}. Experiments were conducted in the Time-of-Flight Medium Energy Ion Scattering (ToF-MEIS) setup in at the Tandem Laboratory \cite{strom2022ion}. An angularly rotatable, position sensitive detector based on micro-channel plates is employed to detect backscattered particles in a total solid angle of 0.13 sr. In this work, only projectiles backscattered within 129° ± 2° are considered for evaluation. All measurements were performed under normal incidence. The beamline is connected to a 350 kV Danfysik ion implanter, used to produce atomic or molecular ion beams with typical energies between 20 and 350 keV. Lower energies are achieved by lowering the extraction potential in the ion source. H$_2^+$ and He$^+$ ions were used with primary energy from 30 to 60 keV and from 10 to 60 keV, respectively, with a corresponding energy resolution ranging from 1.1 to 1.9 keV. 

\subsubsection{Data analysis}

\begin{figure}[t!]
\centering\includegraphics[width=8.7 cm]{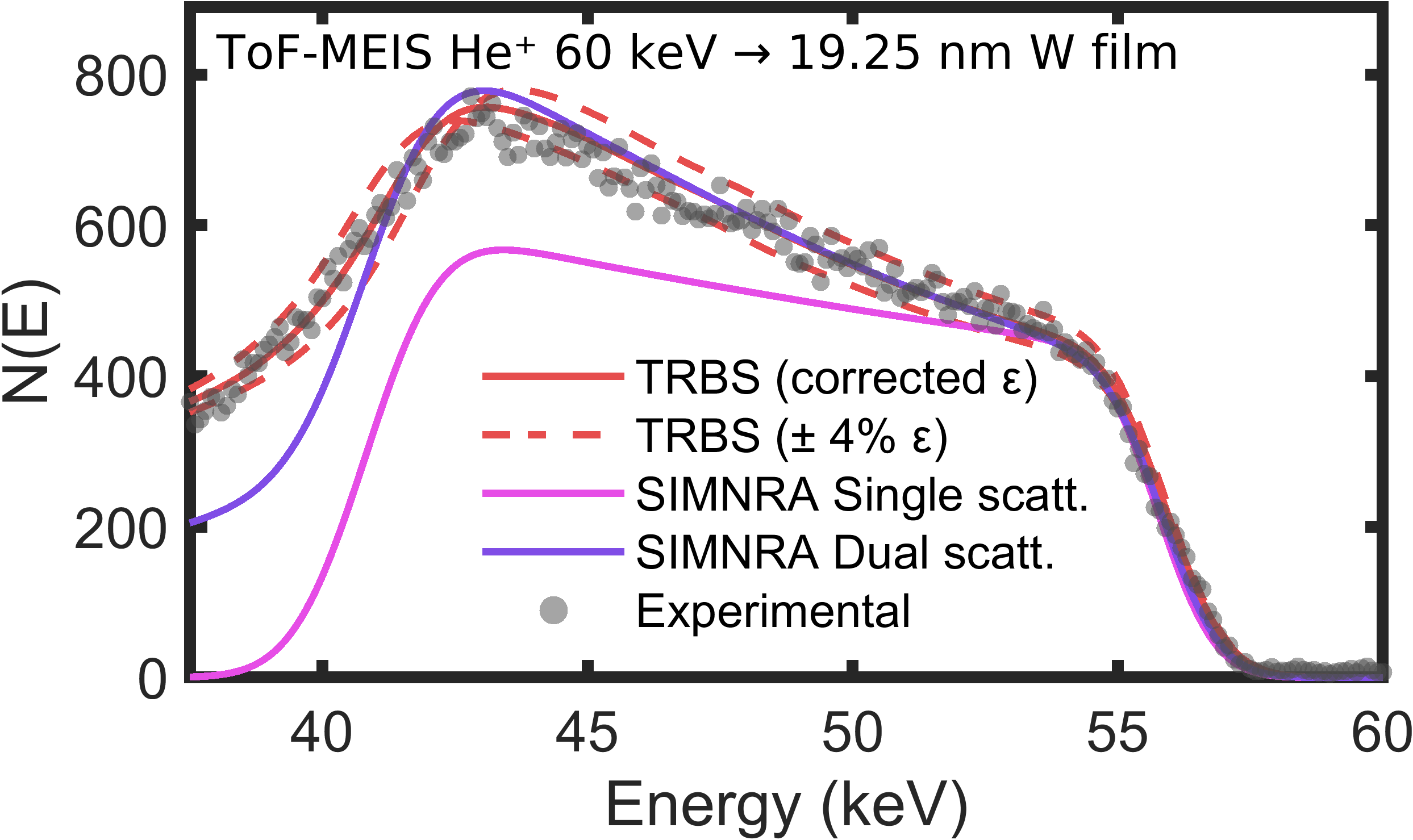}
\caption{Experimental energy converted ToF-MEIS spectrum of 60 keV He$^+$ scattered from a W film on a carbon substrate and respective normalized TRBS simulation with corrected SCS. Dashed lines represent TRBS simulations with ±4 \% of the corrected SCS presented for comparison. SIMNRA simulations are also presented considering both single and multiple scattering. }
\label{fig3}
\end{figure}

To account for multiple and plural scattering, Monte-Carlo (MC) simulations were performed using the TRBS code \cite{biersack1991particularly} with the screened Universal Potential (ZBL). Simulations using TFM potential were also performed, not resulting in significant differences in the simulated backscattering spectra. The stopping cross section (SCS), stopping power normalized by the atomic density of the material, was extracted from the width of the energy converted spectrum. In order to compare energy spectra obtained from the experiment and simulation, the experimental resolution has to be considered. Therefore, simulated data are convoluted with a Gaussian of appropriate width. The energy-converted experimental data and MC simulations were also compared with analytical code SIMNRA \cite{mayer1999simnra} using both single scattering mode (commonly employed in ion beam analysis) or dual scattering with the same corrected SCS from TRBS simulation. An example of experimental spectrum in comparison with TRBS and SIMNRA simulations is presented in Fig. \ref{fig3}. While an excellent agreement between energy-converted experimental data and TRBS with corrected SCS is obtained, single scattering simulation fail to reproduce the multiple scattering background. Significantly better agreement is obtained in dual-scattering mode, with the exception of the lower energy background. This observation further justifies the necessity of employing accurate Monte-Carlo simulations including multiple scattering to precisely extract SCS data.

\section{Results and Discussion}

\subsection{Electronic stopping of hydrogen and helium projectiles in W}

\begin{figure}[b!]
\centering\includegraphics[width=8.7 cm]{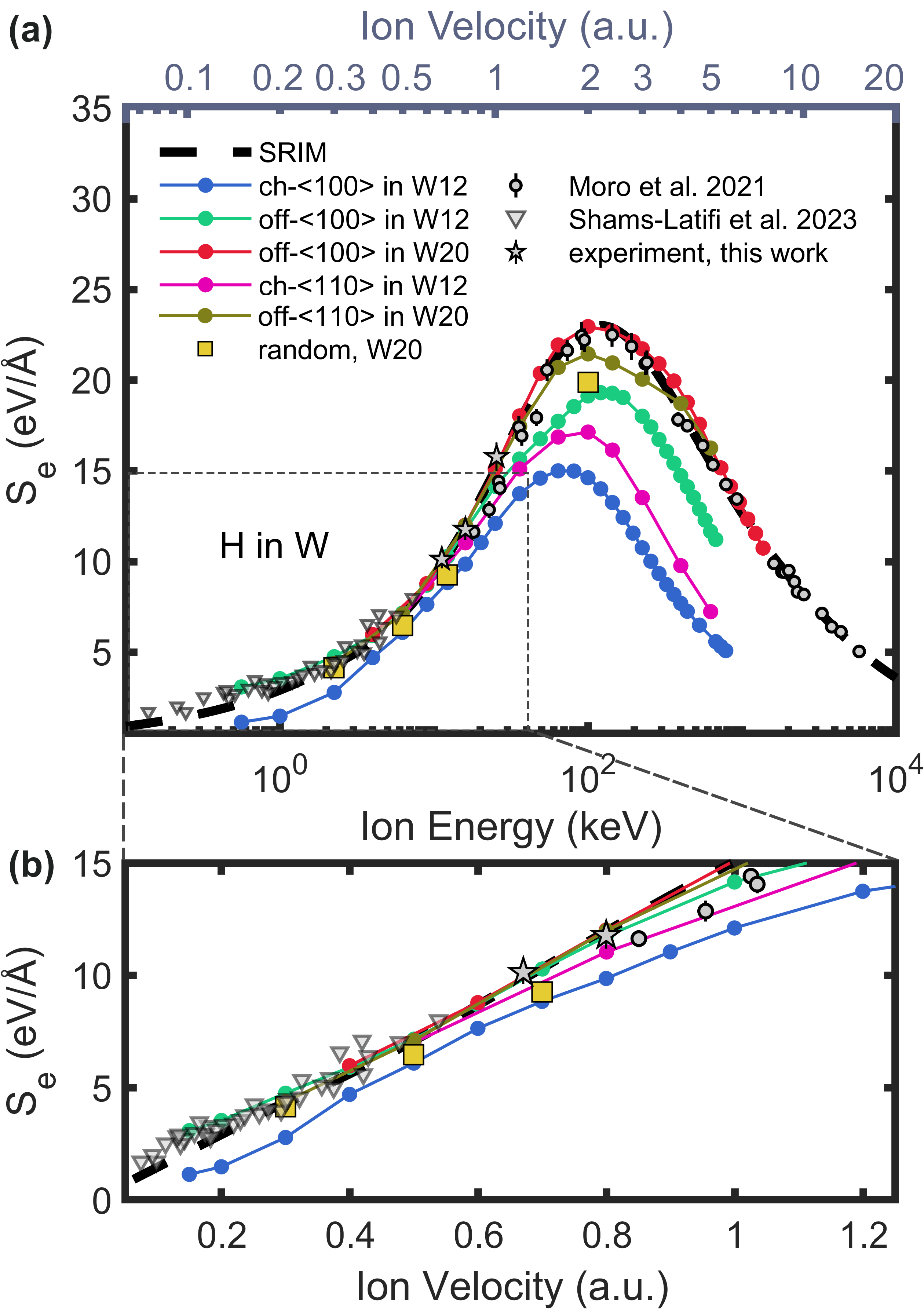}
\caption{(a) Electronic stopping power obtained from TDDFT simulations as a function of energy/velocity for hydrogen exploring different trajectories in tungsten. Simulated data is compared with SRIM-2013 predictions (black dashed line), and experimental data (this work - grey stars, \cite{moro2021experimental} - grey circles, \cite{shams2023experimental} - grey down triangles). (b) A close-up view with a linear velocity-axis scale. } 
\label{fig4}
\end{figure}

\begin{figure}[t!]
\centering\includegraphics[width=8.7 cm]{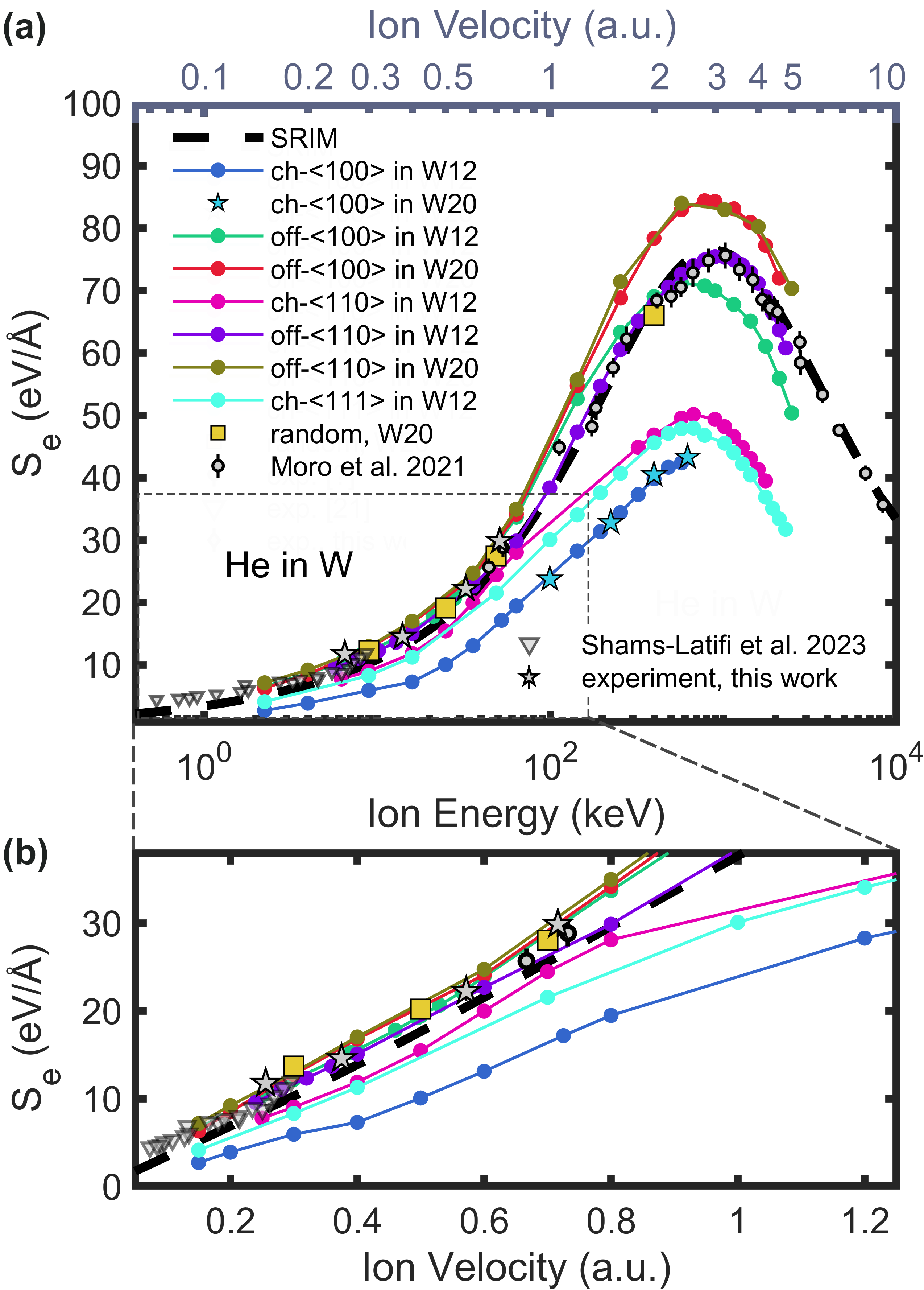}
\caption{(a) Electronic stopping power obtained from TDDFT
simulations as a function of energy/velocity for helium exploring different trajectories in tungsten. Simulated data is compared with SRIM-2013 predictions (black dashed line), and experimental data (this work - grey stars, \cite{moro2021experimental} - grey circles, \cite{shams2023experimental} - grey down triangles). (b) A close-up view with a linear velocity-axis scale. }
\label{fig5}
\end{figure}

We have computed the electronic stopping curves for H and He projectiles across a broad range of velocities (energies) while considering various trajectories within the tungsten lattice. The obtained TDDFT results are presented in Fig. \ref{fig4} and Fig. \ref{fig5}, where a comparison is made with the experimental data (including also data from \cite{moro2021experimental, shams2023experimental}) and predictions from SRIM \cite{ziegler1985stopping}. Although SRIM is a valuable tool, it assumes a homogenous target material and employs semi-empirical models derived from fitting existing experimental data. Consequently, it may not fully capture all the underlying physical mechanisms and its applicability is limited to velocity ranges where experimental data is available. Thus, we can readily note the discrepancies in the low-velocity range, where both TDDFT close approach data and experimental measurements lie above SRIM predictions with a difference of up to 40 \%.

The stopping curves for hyperchanneling trajectories predicted by TDDFT lie below the experimental data for both projectiles. This observation aligns with our findings depicted in Fig. \ref{fig2}a, which clearly demonstrates that the center channeling projectile fails to capture the full electron density distribution in the system. The off-centered channeling directions show a better agreement with the experimental data. In the case of hydrogen, this correspondence is further improved by incorporating semi-core states into the tungsten pseudopotential. As a result, a nearly perfect alignment is achieved between the SRIM predictions, TDDFT data, and experimental observations around the Bragg peak. Conversely, the off-centered helium loses significantly more energy in W20 than is predicted by SRIM around the maximum and over the whole intermediate energy range. These observations suggest that the electronic orbitals of helium play a significant role in energy loss via ion-electron collisions in a high-velocity regime.

Furthermore, Figs. \ref{fig4}, \ref{fig5} reveal variations in the Bragg peak positions among the different datasets. Specifically, using the W12 pseudopotential, there are differences in the stopping maximum position for hydrogen when traveling along $\langle$100$\rangle$ and $\langle$110$\rangle$ channels, as well as for the off-centered position within $\langle$100$\rangle$ (Fig. \ref{fig4}). However, when comparing the green and red curves in the same plot, we observe that the peak stopping power remains consistent when replacing the W12 with the W20 pseudopotential for the same trajectory. These observations lead us to the conclusion that, at least for the H projectile, the Bragg peak position depends on the impact parameter but remains unaffected by a change in pseudopotential. In the case of He projectiles, the results are more intricate, but the impact parameter dependence is still traced in the scenarios involving off-centered $\langle$100$\rangle$ and $\langle$110$\rangle$ channels. These findings imply that both trajectory dependence and the incorporation of inner-shell electrons contribute to the shift in the Bragg peak position. Notably, this contradicts the conclusion reported in~\cite{li2023electronic} that the stopping curve maximum is independent of the impact parameter.

\begin{figure}[t!]
\centering\includegraphics[width=8.7 cm]{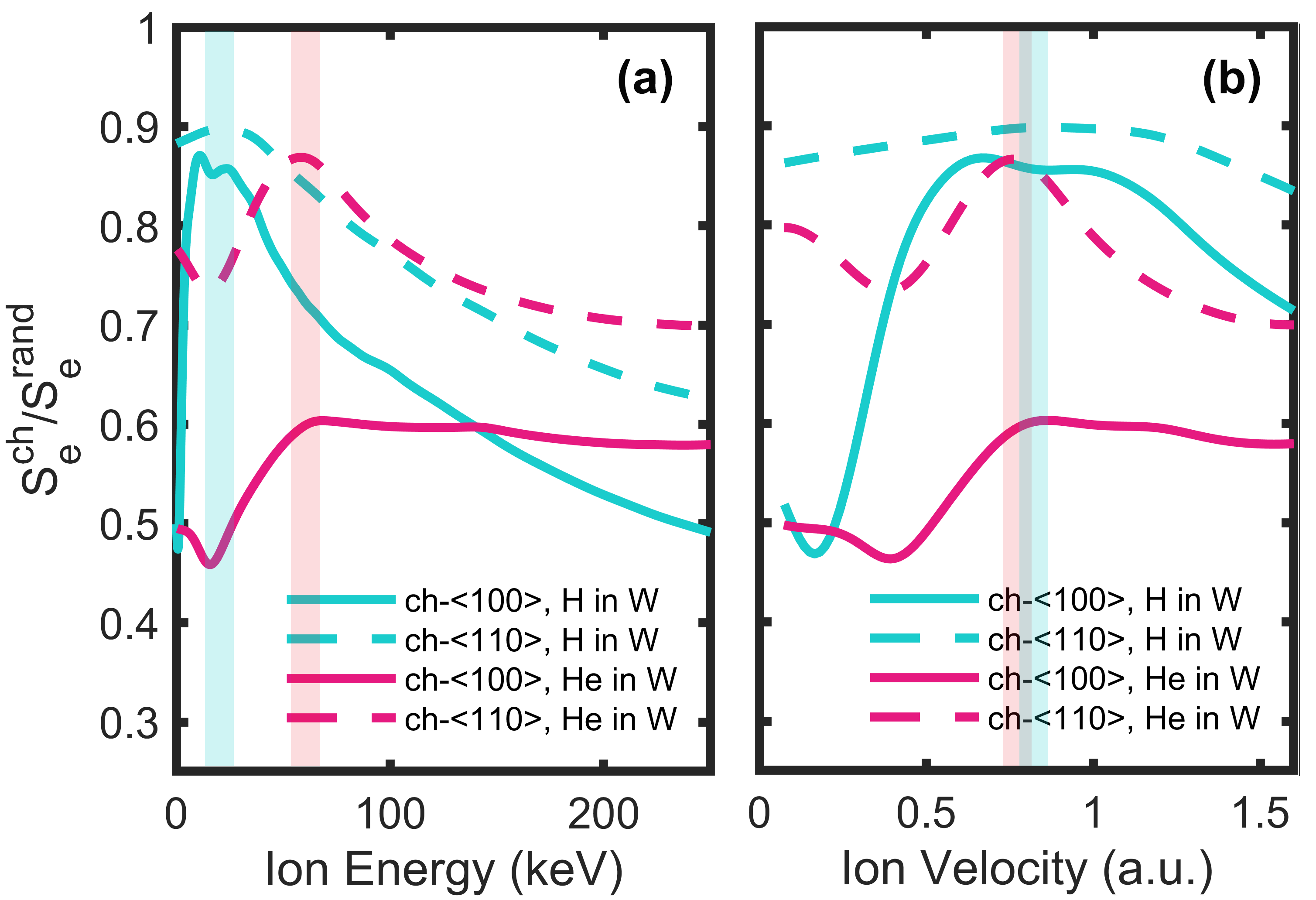}
\caption{Ratio of the electronic energy loss along $\langle$100$\rangle$ and $\langle$110$\rangle$ channeling trajectories to the experimental random data for H (dashed lines) and for He (solid lines) projectiles propagating in W. In (a) the data is plotted relative to ion energy, while in (b) - to its velocity. All the data sets were interpolated; experimental data was smoothed using Savitzky-Golay filter. Red and cyan thick lines are introduced to show how the ratio peak values correspond to each other for two projectiles in the energy and velocity representations. }
\label{fig6}
\end{figure}

Finally, we address the question about the possible differences between the mechanisms through which H and He projectiles lose energy in channeling and close approach trajectories. In the case of hydrogen (Fig. \ref{fig4}a, b), we observe a tendency of the stopping values for different trajectories to converge within the specific velocity range from about 0.4 a.u to 1 a.u. This observation aligns with previous findings for protons in copper and nickel \cite{quashie2016electronic, quashie2021directional}. Interestingly, this trend is not as pronounced for helium projectiles. In \cite{lohmann2020disparate} it was conjectured that at slower particle velocities, the increased interaction time allows for the activation of nonadiabatic processes such as the formation of molecular orbitals and associated electron promotion. This could result in enhanced electronic energy losses during close encounters between helium and target atoms, contrasting with the observed data from hydrogen. To directly compare predictions for H and He in W, in Fig. \ref{fig6} we present a ratio of the electronic stopping power for $\langle$100$\rangle$ and $\langle$110$\rangle$ channeling trajectories to the experimental random data for both projectiles in energy (Fig. \ref{fig6}a) and velocity representations (Fig. \ref{fig6}b). While analysing the same energy interval (from 40 keV to 200 keV) results in a perfect alignment with observations from \cite{lohmann2020disparate}, expanding this range and, moreover, looking at the velocity-dependent picture provides a different outlook. For both projectiles, the ratio for the $\langle$100$\rangle$ channel increases with decreasing ion velocity up to a certain point, after which it drops to then start growing again in the limit of $v \to 0$. Inspection of the data in Fig. \ref{fig4}b and Fig. \ref{fig5}b reveals that this behavior is shaped by a deviation of the electronic stopping power from a linear velocity dependence along a hyperchanneling trajectory. The nature of this deviation will be discussed in the following section.

\subsection{Nonlinear velocity dependence for hyperchanneling projectiles}

Both hydrogen and helium electronic stopping data for the hyperchanneling directions exhibit nonlinear behaviour in the low-velocity range (see the blue and pink curves in Fig. \ref{fig4}b and Fig. \ref{fig5}b). Previously, similar findings were reported in a number of experimental and TDDFT-based works for different metals \cite{cantero2009velocity, primetzhofer2011electronic, quashie2016electronic, li2021first, li2023electronic2}. In \cite{primetzhofer2011electronic}, such nonlinearity (observed only for He) was explained through charge exchange processes occurring between a helium projectile and the target atom. However, since these processes are distance-dependent, it contradicts the observation of this effect in hyperchanneling trajectories, a trend commonly observed in theoretical works (e.g., \cite{quashie2016electronic, li2021first, li2023electronic2}). Predictions from the free electron gas (FEG) model, however, led to the conclusion that band and bound effects contribute to the emergence of slopes in the velocity curve for both hydrogen and helium projectiles.

\begin{figure*}[t!]
\centering\includegraphics[width=\linewidth]{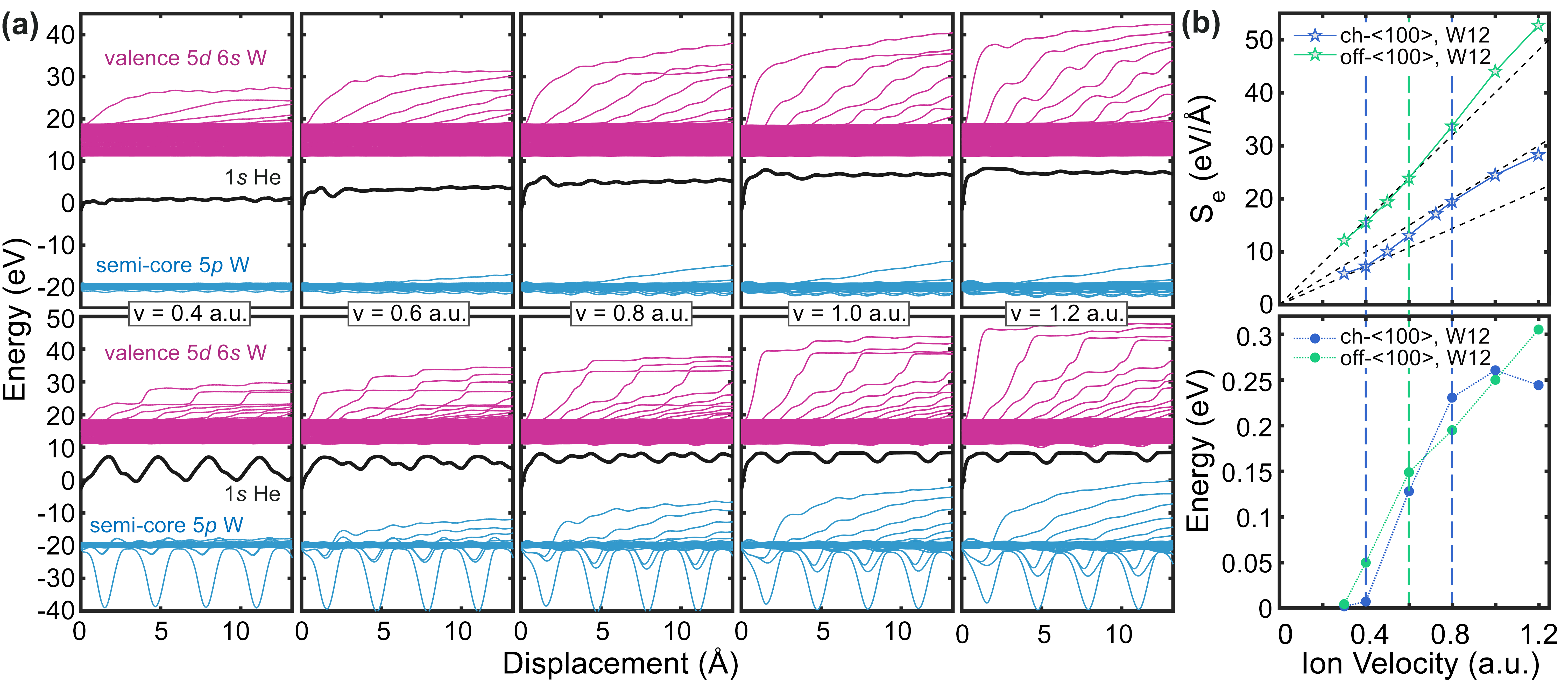}
\caption{Analysis of the nonlinear velocity dependence of the electronic stopping. (a) Instantaneous energy expectation values for the propagating Kohn-Sham states via the projectile displacement. The upper row depicts results for $\langle$100$\rangle$ hyperchanneling trajectory, while the bottom row illustrates those for $\langle$100$\rangle$ off-centered channel in W12. The time evolution of the orbitals energy is presented for velocities ranging from 0.4 to 1.2 a.u. (b) Comparison between the velocity dependence of the electronic stopping power and the rate of change in the highest 5$p$-state energy expectation values for the two examined trajectories. The dashed blue and green lines illustrate a correspondence in characteristic regions with a slope change for both dependences. }
\label{fig7}
\end{figure*}

In the following, we analyze the observed deviation from a linear velocity dependence for He hyperchanneling in W in more detail. Fig. \ref{fig7}a presents the energy expectation values for both propagating valence and semicore states at the projectile velocities where the change in slope occurs. These results are compared with the off-centered channeling He in W12, depicted in the bottom row of Fig. \ref{fig7}a. The analysis reveals that activation of the inner-shell 5$p$-electrons occurs exclusively when the velocity surpasses 0.4 a.u. Meanwhile, no visible difference in the dynamics is observed between the cases where $v$ = 0.8 a.u. and $v$ = 1.2 a.u., suggesting that further excitation of these states remains constrained at higher velocities. Conversely, the overall picture for He moving in closer proximity to the target atoms is different. Here, we observe, that at $v$ = 0.4 a.u., the 5$p$ states are already activated, and their expectation energy values gradually increase with velocity.

To directly compare the dynamics of expectation energy values and electronic energy losses, we extract a linear slope from the highest semicore state, avoiding the transient region, and plot it via velocity in Fig. \ref{fig7}b. The dashed lines identify the characteristic regions seen in both plots. Thus, the kink from 0.4 a.u. to 0.8 a.u. for the hyperchanneling data (blue) is clearly visible in the bottom plot, confirming this to be a result of the semicore electron states activation. Similarly, a change in slope found in the electronic stopping curve for off-center channeling He (green) corresponds to an analogous feature in the semicore states expectation energy data. In this case, however, the region from $v$ = 0.6 a.u. onwards could indicate the initiation of nonlinear plasmalike behavior.

\subsection{Environment dependence of electronic stopping}

\begin{figure}[t!]
\centering\includegraphics[width=8.7 cm]{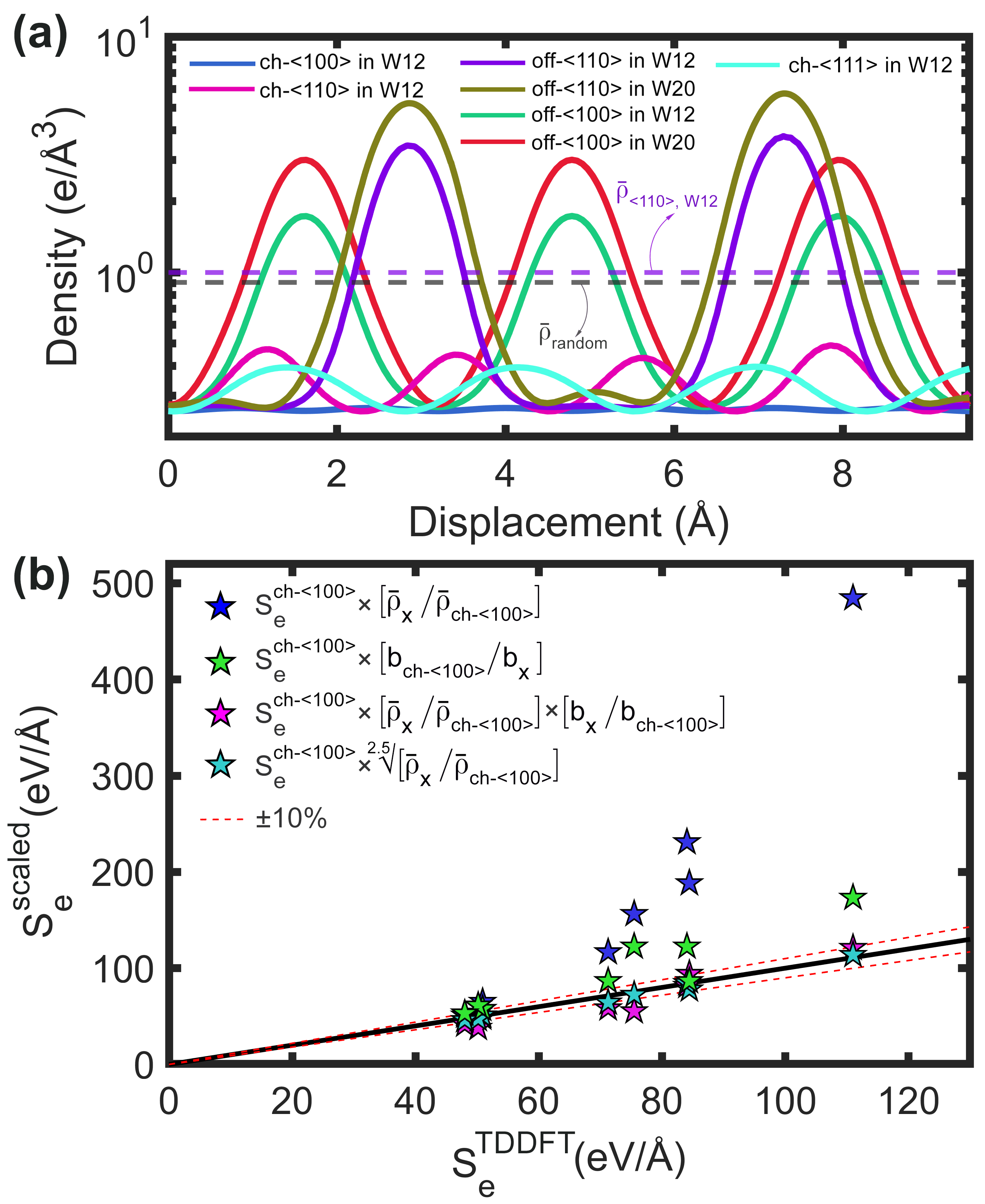}
\caption{Environment dependence of the electronic stopping. (a) Electron densities along the different trajectories in W12 and W20. The choice of trajectories corresponds to Fig. \ref{fig5}. A logarithmic scale is used for better visibility of the low-amplitude densities for hyperchanneling directions. The dashed violet and grey lines illustrate the mean density value for the $\langle$110$\rangle$ off-centered channeling trajectory in W12 and random trajectories, respectively. (b) Electronic stopping power calculated by employing scaling coefficients for various trajectories traveled by He in W12 and W20 with respect to the TDDFT data. Black solid line corresponds to the identity function. Dashed red lines define $\pm10\%$ deviation from the TDDFT predictions.}
\label{fig8}
\end{figure}

A prominent feature in Fig. \ref{fig5} is the excellent alignment of the data for off-centered He channeling in $\langle$110$\rangle$ of W12 with the experimental results. This alignment suggests that the density sampled along this trajectory could be be seen as representative of the density experienced by ions in the experiment, and ideally, through randomly implemented trajectories. Upon revisiting Fig. \ref{fig2}a, we observe that the $\langle$110$\rangle$ trajectory in W12 effectively matches the reference electron density distribution, confirming the earlier assumption. Likewise, comparing the mean electron density value for this channel with the average density determined using pre-sampled random trajectories yields estimations of 1.0 e$/\text{\angstrom}^3$ and 0.91 e$/\text{\angstrom}^3$, respectively (see the dashed violet and grey lines in Fig. \ref{fig8}a). The random data slightly underestimates measurements around the Bragg peak, and this is also reflected in the ratio of the mean density values. All these findings allow making a following assumption: the electron density plays a pivotal role in determining the stopping power in that region.

To gain deeper insights into the relationship between electronic stopping and its environmental factors, we explore a potential link between energy loss and both electron density and impact parameters. Fig. \ref{fig8}a shows density values calculated along all the trajectories depicted in Fig. \ref{fig5} for He propagating in tungsten. Some correlation is apparent; for instance, both the mean density and the electronic stopping values are very close to each other for the case of $\langle$110$\rangle$ and $\langle$111$\rangle$ channels. However, comparing the $\langle$100$\rangle$ channel for W12 and W20 (green and red curves in Fig. \ref{fig5} and Fig. \ref{fig8}a, respectively), we find that while the mean density value is 1.6 times larger for W20, the stopping values reveal noticeably lower proportionality of 1.2 (this holds for hydrogen). Additionally, we notice that while the average electron density is 1.2 times higher for the $\langle$110$\rangle$ off-centered trajectory in W20, the resulting electronic stopping curve overlaps with that for helium in the $\langle$100$\rangle$ channel in W20 (olive green and red curves, respectively).

We further expand this analysis by examining if scaling the data for one of the channels using a simple coefficient could match the TDDFT calculations for any other trajectory. Thus, in Fig. \ref{fig8}b we present the results obtained from scaling the $\langle$100$\rangle$ channel data using coefficients related to density and impact parameters for the trajectories shown in Fig. \ref{fig5}. We additionally analyze the data for a He projectile shifted by 1/8 $a_0$ and 3/8 $a_0$ from the central position in the $\langle$100$\rangle$ channel in W20. We explored the following coefficients: (1) the ratio of mean electron densities $\bar{\rho}_\text{x}/\bar{\rho}_{\text{ch-}\langle100\rangle}$, (2) the ratio of impact parameters $b_{\text{ch-}\langle100\rangle}/b_\text{x}$, (3) a combination of density and impact parameter ratios $\bar{\rho}_\text{x}/\bar{\rho}_{\text{ch-}\langle100\rangle} \times b_\text{x}/b_{\text{ch-}\langle100\rangle}$, and (4) the nth root (where n = 2.5) of the mean electron density ratio $\sqrt[2.5]{\bar{\rho}_\text{x}/\bar{\rho}_{\text{ch-}\langle100\rangle}}$, where $\text{x}$ represents any other trajectory of interest. 

Fig. \ref{fig8}b clearly demonstrates that scaling the data for $\langle$100$\rangle$ center channel position with the linear electron density ratio (blue star symbols) fails to accurately predict the electronic stopping power for other scenarios. The error grows quite fast when we transition from examining the hyperchanneling to the close approach trajectories. Depicted by the green color, scaling with the impact parameter ratio appears to offer a somewhat improved perspective, although still falling short, resulting in a deviation of up to 50 $\%$. A further attempt to balance the contributions from these parameters by dividing the coefficient (1) by (2) (pink star symbols) yields better predictions, however, fluctuating rather unpredictably across trajectories (for example, surpassing 10 $\%$ deviation for the hyperchanneling directions). Ultimately, we discover that considering solely the difference in the sampled density across various channels, albeit in a nonlinear fashion, yields the most accurate predictions (illustrated by cyan star symbols). This result validates our initial assumption regarding the electron density pivotal role in determining the maximum electronic stopping power. However, understanding the underlying nature of this dependency necessitates further, more comprehensive investigation.

\section{CONCLUSIONS}

In conclusion, we have reported the electronic stopping power of hydrogen and helium projectiles in tungsten across a wide range of velocities, focusing particularly on the low-velocity range where reliable database results are lacking. By employing a pre-sampling algorithm to search for optimal random trajectories within the TDDFT framework, we achieved good agreement between experimental and theoretical stopping curves. In the low-velocity range, we found a linear behavior in random and close approach electronic stopping data, although, for hyperchanneling trajectories, we identified distinct slopes. This phenomenon was attributed to the activation of semicore states once the projectile velocity surpasses a certain threshold. Additionally, we observed a significant difference in how the inclusion of semicore states to the tungsten pseudopotential affects H and He stopping around the Bragg peak. While substituting the W12 pseudopotential with W20 notably enhanced agreement with experimental data for helium projectiles, it did not yield a similar improvement for hydrogen, maintaining an underestimation of the averaged electronic stopping. Finally, our study revealed a pronounced environment dependence of high-energy electronic stopping, that was shown to be well-defined by a nonlinear function of the electron density.

The synergistic approach explored here, which combines ab initio predictions and experimental measurements, serves as a reliable tool for studying electronic energy losses of ions in different materials. Beyond its fundamental significance, the electronic stopping data will provide essential input for larger-scale atomistic simulations, facilitating comprehensive analysis of ion-irradiation effects. Enhanced damage prediction accuracy holds crucial importance for materials science, nuclear engineering, medicine, and semiconductor technology.

\section*{ACKNOWLEDGMENTS}

This work was carried out on the EUROfusion High Performance Computing facility Marconi-Fusion hosted at Cineca. Support of the accelerator operation at Uppsala University by VR-RFI (contract \# 2019-00191) is gratefully acknowledged. This work has been carried out within the framework of the EUROfusion Consortium, funded by the European Union via the Euratom Research and Training Programme (Grant Agreement No 101052200 — EUROfusion). Views and opinions expressed are however those of the author(s) only and do not necessarily reflect those of the European Union or the European Commission. Neither the European Union nor the European Commission can be held responsible for them.


\bibliography{apssamp}

\end{document}